\documentclass[aps,pra,showpacs,twocolumn,superscriptaddress]{revtex4-1}
\usepackage{amsmath}
\usepackage{amssymb}
\usepackage{graphicx}
\usepackage{epstopdf}
\usepackage{times,txfonts}
\usepackage{easybmat}
\usepackage{mathtools}

\usepackage[bookmarks=false]{hyperref}
\hypersetup{colorlinks=true,citecolor=blue,linkcolor=blue,urlcolor=blue,pdfstartview=FitH,bookmarksopen=true}
\usepackage{color,soul}

\newtheorem{theorem}{Theorem}
\newtheorem{lemma}{Lemma}

\begin{document}
\title{Local distinguishability of quantum states in bipartite systems}
\author{Xiaoqian Zhang}
\thanks{These authors contributed equally to this work.}
\affiliation{College of Information Science and Technology, Jinan University, Guangzhou 510632, China}
\author{Cheng Guo}
\thanks{These authors contributed equally to this work.}
\affiliation{Institute for Advanced Study, Tsinghua University, Beijing 100084, China}
\author{Weiqi Luo}
\email{lwq@jnu.edu.cn}
\affiliation{College of Information Science and Technology, Jinan University, Guangzhou 510632, China}
\author{Xiaoqing Tan}
\affiliation{College of Information Science and Technology, Jinan University, Guangzhou 510632, China}

\date{\today}
\pacs{03.67.Lx, 03.67.Dd, 03.65.Ud.}

\begin{abstract}
  In this article, we show a sufficient and necessary condition for locally distinguishable bipartite states via one-way local operations and classical communication (LOCC). With this condition, we present some minimal structures of one-way LOCC indistinguishable quantum state sets. As long as an indistinguishable subset exists in a state set, the set is not distinguishable. We also list several distinguishable sets as instances.
\end{abstract}

\maketitle

\section{Introduction}
Quantum entanglement is an important manifestation of quantum nonlocality \cite{Einstein1935}. Quantum entangled states, especially maximally entangled states, play an important role in quantum information theory \cite{Ekert91,Bennett1992,Bennett1993,Barenco1996}. Maximally entangled states have attracted considerable attention in recent years, since they can assist us in managing quantum information and understanding the fundamental principles of quantum mechanics. A given set of quantum states is distinguishable, if we can identify each state in this set unambiguously. Otherwise, this set is so-called indistinguishable. Many scholars have studied the distinguishable \cite{Tian2015,Cao2004,Ghosh2004,Zhang015,Zhang5,Bandyopa11} and the indistinguishable \cite{Zhang014,Wang2016,Yu12,Li2015,Yu2015} maximally entangled states. In \cite{Bandyopa11}, Bandyopadhyay \emph{et al.} revealed an important conclusion: unilaterally transformable quantum states are distinguishable by one-way local operations and classical communication (LOCC). With this conclusion, they proved that maximally entangled states are exactly indistinguishable by one-way LOCC in $C^d\otimes C^d$ ($d=4, 5, 6$). Zhang \emph{et al.} pointed out two indistinguishable classes of maximally entangled states by one-way LOCC in $C^d\otimes C^d$ ($d \leq 10$) \cite{Zhang014}. The definitions of one-way LOCC and two-way LOCC are given as follows.

The different subsystems of quantum states may be in distance, so the following two restrictions are reasonable: (1) (local operations) each subsystem can only perform operations on its own but global operations on the composite system are not allowed, and (2) (classical communication) each party can transmit other parties its measurement outcomes by classical channels in order to coordinate and communicate \cite{Cohen07}. The protocols with these two rules are called ``local operations and classical communication" protocols, abbreviated as LOCC. The LOCC protocols can be further refined based on the use of classical communication. In one-way LOCC, the classical communication is only transmitted from one part (Alice) to the other (Bob), but no information is allowed to move in the other direction (Bob to Alice). For two-way LOCC, Alice and Bob can communicate enough rounds after each turn of his or her local operations. In the following texts, the LOCC means two-way LOCC conveniently.

Quantum nonlocality of orthogonal product states was widely studied, which included the local distinguishability \cite{Duan09,Bandy2012,Yang2013} and the local indistinguishability \cite{Duan2010,Feng09,Zhang14,Wang15,Zhang15,Zhan15}. Duan \emph{et al.} exhibited the distinguishable orthogonal product states by separable operations in $C^3\otimes C^3$ and $C^2\otimes C^2\otimes C^2$ dimensional quantum systems, respectively \cite{Duan09}. However, this result is not always true in LOCC protocols because LOCC operations are weaker than separable operations. These orthogonal product states are all indistinguishable by LOCC, which exhibits quantum nonlocality without entanglement in \cite{Duan2010,Feng09,Zhang14,Wang15,Zhang15,Zhan15}. Many other interesting works are presented about quantum distinguishable problems in \cite{Walgate00,Band09,Duan07,Duan08,feng07,duan11,duan14,Nath05,Chen03,Singal2016,Nathanson2013,Horo2003}. Singal proposed a framework for distinguishing orthogonal bipartite states by one-way LOCC \cite{Singal2016}. Similarly, Nathanson showed that any set of three orthogonal maximally entangled states can be distinguished via one-way LOCC with high probability \cite{Nathanson2013}.

The distinguishability of complete orthogonal product states can be proved by using the sufficient and necessary conditions in \cite{Chen2004,Ma2014}. Walgate and Hardy also researched this problem for distinguishing bipartite orthogonal quantum states in $C^2\otimes C^n$ by LOCC \cite{Horod03}. The sufficient and necessary condition was used to prove the local distinguishability (or indistinguishability) of a class orthogonal product states. Zhang \emph{et al.} extended and proved the indistinguishability of orthogonal product states in $C^d\otimes C^d$ ($d$ is odd) \cite{Zhang14}. Not only the distinguishability of pure states has already been researched widely but also 
mixed states \cite{Duan2010,Herzog04,Feng04,Zhang06,Stojn07}. Duan \emph{et al.} proved the distinguishability of mixed states in $C^2\otimes C^n$ \cite{Duan2010}. Feng \emph{et al.} pointed out that if and only if the states from a mixed state set are orthogonal, these states can be unambiguously discriminated \cite{Feng04}.

In this paper, we focus on studying local distinguishability of bipartite quantum state sets. In our protocol, Alice is the first person to perform a nontrivial measurement. We first present the sufficient and necessary condition for locally distinguishable bipartite quantum states in Sec. \ref{sec:LOCC}. Secondly, this condition for distinguishability of a certain state set can be used to prove the local indistinguishability of orthogonal product states in $C^{(3l_A+1)} \otimes C^{(3l_B+1)}$ ($1\leq l_A\leq l_B$). We also analyse and prove the distinguishability of three and four orthogonal quantum states via one-way LOCC in Sec. \ref{sec:jud}. We assess the number of product states in a $C^{d_A}\otimes C^{d_B}$ distinguishable set via one-way LOCC when Alice goes first. Furthermore, we analyze and show the minimum structures of one-way LOCC indistinguishable orthogonal product states in Sec. \ref{sec:one}.  Finally, the conclusions are shown in Sec. \ref{sec:con}.

\section{A sufficient and necessary condition for bipartite distinguishable states }
\label{sec:LOCC}
When global operations are allowed, a state set is distinguishable if and only if these quantum states in the set are orthogonal to each other.
\begin{lemma}\label{A4}
A given set of quantum states $\{ \rho^{1}, \rho^{2}, \cdots, \rho^{N}\}$ are distinguishable
if and only if
$\forall h\neq k,~ \text{tr}(\rho^h\rho^k)=0.$
\end{lemma}

By the way, when the set $\{ \rho^{1}, \rho^{2}, \cdots, \rho^{N} \}$ is a pure state set $\{|\psi_1\rangle, |\psi_2\rangle, \cdots, |\psi_N\rangle\}$, the condition $\text{tr}(\rho^h\rho^k)=0$ can be rewritten as $ \langle \psi_h|\psi_k\rangle=0.$

After introducing the definitions of one-way LOCC and LOCC in introduction, we will introduce an important lemma (Lemma \ref{A2}) \cite{Singal2016}, by which one can simplify the measure to an orthogonality preserving rank-one positive-operator valued measure (POVM) in the one-way LOCC protocols, which checks the distinguishability of pure and mixed states.

\begin{lemma}
\cite{Singal2016,Nathanson2013}\label{A2} Alice can commence a one-way LOCC protocol to distinguish among $\rho^{1}_{AB}, \cdots ,\rho^{N}_{AB}$ if and only if there exists a protocol which starts with an orthogonality preserving rank-one POVM on the side of Alice.
\end{lemma}

Now, one of our main results can be introduced: the sufficient and necessary condition of the bipartite distinguishable quantum state sets via one-way LOCC. Notice, a rank-one POVM can be presented via some certain basis.

\begin{theorem}
Alice and Bob share a $C^{d_A}\otimes C^{d_B}$ composite system. For any set $S = \{\rho_1, \rho_2, \cdots, \rho_N\}$ in this system, $S$ is one-way LOCC distinguishable when Alice goes first, if and only if there exists a basis $\{|1\rangle, |2\rangle, \cdots, |d_A\rangle\}_A$, such that for any element $\rho_x$,
\begin{eqnarray}
\rho_x=\sum\limits_{m,n=1}^{d_A}|m\rangle\langle n|\otimes\rho_{mn}^x, \ ~(x=1, 2, \ldots, N),
\end{eqnarray}
where $\{\rho^x_{mm}\}$ is a set of orthogonal non-normalized states, may be zero sometimes, $(m = 1, \ldots, d_A ~ ),$ and following condition holds:
\begin{eqnarray}
\forall h\neq k,~ \text{tr}(\rho^h_{mm}\rho^k_{mm})=0, \ ~(m = 1, \ldots, d_A).
\end{eqnarray}
\end{theorem}

\emph{Proof:}
The sufficiency is obvious. For the necessity, by \emph{Lemma \ref{A2}}, when Alice commences a one-way LOCC protocol, the basis $\{|1\rangle, |2\rangle, \cdots, |d_A\rangle\}_A$ is according to her orthogonality preserving rank-one POVM. The reason of ``$\forall h \neq k, \text{tr}(\rho^h_{mm}\rho^k_{mm})=0 $'' is the distinguishability of states in Bob's subsystems. Otherwise, $\rho_h$ and $\rho_k$ are indistinguishable. By \emph{Lemma \ref{A2}}, if the state set is indistinguishable via all rank-one POVM elements, the set is indistinguishable for all one-way LOCC protocols. The proof is completed. $\square$

When $S$ is a pure state set, we can obtain the following theorem.
\begin{theorem}\label{A3}
A $C^{d_A} \otimes C^{d_B}$ pure quantum state system is shared by Alice and Bob. For any set $S$ in this system, $S$ is one-way LOCC distinguishable when Alice goes first, if and only if there exists a basis $\{|1\rangle, |2\rangle, \cdots, |d_A\rangle\}_A$, such that for any element $|\psi_k\rangle$ in $S$,
\begin{eqnarray}\label{B2}
|\psi_k\rangle=\sum\limits_{j=1}^{d_A} |j\rangle_A|\eta_j^k\rangle_B, ~\text{and}~
\forall h \neq k, \langle\eta_j^h|\eta_j^k\rangle=0.
\end{eqnarray}
\end{theorem}

When $d_A =2$, these two theorems will be the result shown by Walgate and Hardy \cite{Horod03} and the result from Feng, Duan and Ying \cite{Duan2010}.

\section{Bipartite distinguishable pure quantum states in $C^{d_A}\otimes C^{d_B}$ system}
\label{sec:jud}
\begin{figure}[htp]
\centering
\includegraphics[scale=0.5]{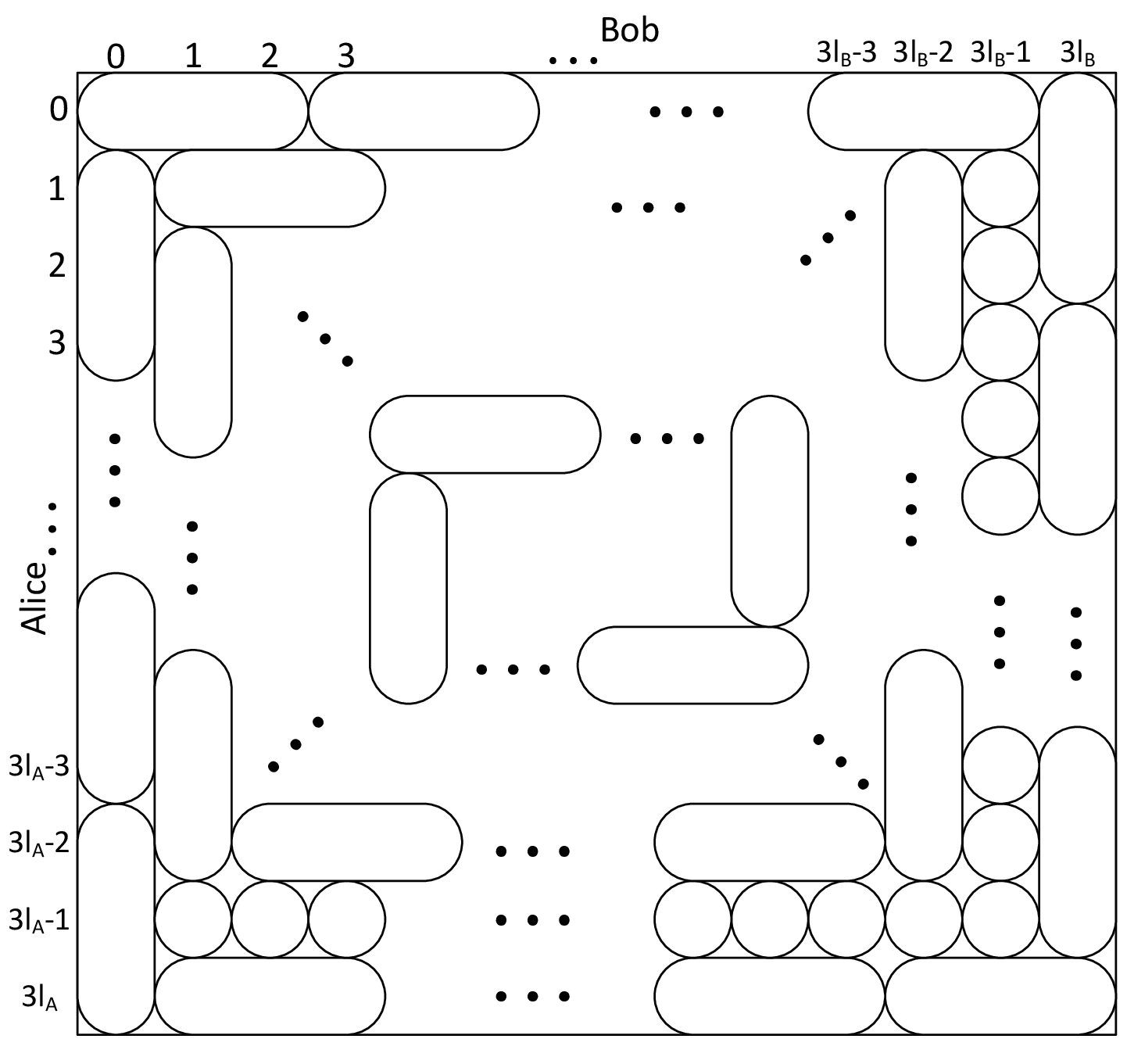}
\caption{The tiling structure of $C^{(3l_A+1)}\otimes C^{(3l_B+1)}$ dimension orthogonal product states $(1\leq l_A\leq l_B)$.}\label{A1}
\end{figure}

Theorem \ref{A3} shows a sufficient and necessary condition for distinguishable high-dimensional state sets with the concrete forms. It is a useful tool for researchers to analyze whether concrete state sets are one-way local distinguishable or not, such as these following orthogonal product states in Eq.(\ref{B1}) (also FIG. \ref{A1}).

\begin{eqnarray}\label{B1}
\begin{array}{l}
\displaystyle |\varphi_{a_tb_t}^\pm\rangle=|a_t\rangle_A(\frac{1}{2}|b_t\rangle\pm\frac{1}{\sqrt{2}}|(b_t+1)\rangle+\frac{1}{2}|(b_t+2)\rangle)_B,\\ \displaystyle \qquad where \ t=1, 2\ \text{and}  \\
\displaystyle \qquad a_1=0,1, 2, 3, \cdots, l_A-1, \\
\displaystyle \qquad b_1=a_1, a_1+3, a_1+6, \cdots,3l_B-2a_1-3.\\
\displaystyle \qquad a_2=3l_A, 3l_A-2, 3l_A-4, \cdots, 3l_A-2(l_A-1),\\
\displaystyle \qquad b_2=(3l_A-a_2)/2+1, (3l_A-a_2)/2+4, (3l_A-a_2)/2+7,\\
\displaystyle \qquad\qquad \cdots,3l_B-(3l_A-a_2)-2.\\
\displaystyle |\varphi_{a_rb_r}^\pm\rangle=(\frac{1}{2}|a_r\rangle\pm\frac{1}{\sqrt{2}}|(a_r+1)\rangle+\frac{1}{2}|(a_r+2)\rangle)_A|b_r\rangle_B, \\ \displaystyle \qquad where \ r=1, 2 \ \text{and}\ \\
\displaystyle \qquad b_1=0, 1, 2, 3, \cdots, l_A-1,\\
\displaystyle \qquad a_1=b_1+1, b_1+4, b_1+7, \cdots,3l_A-2b_1-2. \\
\displaystyle \qquad b_2=3l_B, 3l_B-2, 3l_B-4, \cdots, 3l_B-2(l_A-1),\\
\displaystyle \qquad a_2=(3l_B-b_2)/2, (3l_B-b_2)/2+3, (3l_B-b_2)/2+6,\\
\displaystyle \qquad \qquad \cdots, 3l_A-(3l_B-b_2)-3.\\
\displaystyle |\varphi_{q_1c_1}\rangle=|(3l_A-1)\rangle_A|c_1\rangle_B, where\  c_1=1, 2, 3, \cdots, 3l_B-1.\\
\displaystyle |\varphi_{q_2c_2}\rangle=|q_2\rangle_A|(3l_B-1)\rangle_B, where \ q_2=1, 2, 3, \cdots, 3l_A-2.
\end{array}
\end{eqnarray}

According to Theorem \ref{A3}, we have the following conclusions.

\vspace{2mm}
{\bf Corollary 1}
\emph{In $C^{3l_A+1}\otimes C^{3l_B+1}$ $(1\leq l_A\leq l_B)$ quantum system, there exists an indistinguishable set via one-way LOCC no matter who goes first, which contains $4l_Al_B+7l_A+3l_B-3$ orthogonal product states $|\varphi_s\rangle$ (in Eq.(\ref{B1})).}

By Theorem \ref{A2}, Corollary 1 can be proved (details in Appendix). Note that in Corollary 1, the indistinguishability of the set in Eq. (\ref{B1}) can be directly proved via our Theorem \ref{A3}, since we can consider the subspaces $C^3\otimes C^{(3l_B+1)}$ in this set. Then extend it to the general case: the dimension of Alice's subsystems is greater than three, our Theorem \ref{A3} is also applicable.
\\

Next, let us check how many orthogonal product states in a state set are always distinguishable via one-way LOCC. Walgate \emph{et al.} pointed out that any three orthogonal product states are always distinguishable via LOCC in $C^2 \otimes C^2$ system \cite{Horod03}. Now, we show the general case.

{\bf Corollary 2}
\emph{Any three orthogonal product states are always distinguishable via one-way LOCC when Alice commences.}

\emph{Proof:}
Suppose three orthogonal product states are
$$|\psi_0\rangle=|a_0b_0\rangle_{AB},\ |\psi_1\rangle=|a_1b_1\rangle_{AB},\ |\psi_2\rangle=|a_2b_2\rangle_{AB}.$$

If $\langle a_0|a_1\rangle=\langle a_0|a_2\rangle=\langle a_1|a_2\rangle = 0,$ it is obvious that Formula (\ref{B2}) holds.

If two of $|a_0\rangle, |a_1\rangle$ and $|a_2\rangle$ are not orthogonal, Formula (\ref{B2}) also holds. We give the proof as follows:

Without loss of generality, suppose $\langle a_1|a_2\rangle\neq 0$ and $|\psi_0\rangle=|a_0b_0\rangle_{AB}=|00\rangle_{AB},$ we have $\langle b_1|b_2\rangle=0.$ There are $4$ different cases:
$1)$ $ \langle 0|a_1\rangle=\langle 0|a_2\rangle = 0,$~
$2)$ $ \langle 0|a_1\rangle=0$ but $\langle 0|a_2\rangle\neq 0,$~
$\\3)$ $ \langle 0|a_2\rangle=0$ but $ \langle 0|a_1\rangle\neq 0,$
$4)$ $\langle 0|a_1\rangle\neq 0$ and $\langle 0|a_2\rangle\neq 0.$

For case $1)$, a basis $\{|0\rangle, |a_1\rangle, |a'_2\rangle\}$ can be constructed by Gram-Schimdt orthogonalization, where the state $|a'_2\rangle=(|a_2\rangle-\alpha|a_1\rangle)$ can be constructed from $\{|0\rangle, |a_1\rangle\}_A$, thus we have
\begin{eqnarray*}
\begin{array}{l}
\displaystyle |\psi_0\rangle=|00\rangle_{AB}, \ |\psi_1\rangle=|a_1b_1\rangle_{AB},\\
\displaystyle |\psi_2\rangle=\alpha |a_1b_2\rangle_{AB}+(|a_2\rangle-\alpha|a_1\rangle)_A|b_2\rangle_B.
\end{array}
\end{eqnarray*}
It means that Formula $(\ref{B2})$ holds. Similarly, the two elements $\{|0\rangle, |a_2\rangle\}_A$ can also construct a basis $\{|0\rangle, |a'_1\rangle, |a_2\rangle\}$ by Gram-Schimdt orthogonalization. So we have
\begin{eqnarray*}
\begin{array}{l}
\displaystyle |\psi_0\rangle=|00\rangle_{AB},\ |\psi_2\rangle= |a_2 b_2\rangle_{AB}, \\ \displaystyle|\psi_1\rangle=\alpha|a_2b_1\rangle_{AB}+(|a_1\rangle-\beta|a_2\rangle)_A|b_1\rangle_B,
\end{array}
\end{eqnarray*}
where $|a'_1\rangle=(|a_1\rangle-\beta|a_2\rangle).$ This is a form as Formula $(\ref{B2}).$

For case $2)$, we get the same basis $\{|0\rangle, |a_1\rangle, |a'_2\rangle\}_A$ as the case 1).

For case $3)$, we get the same basis $\{|0\rangle, |a'_1\rangle, |a_2\rangle\}_A$ as the case 1).

For case $4)$, it means $\langle b_0|b_1\rangle = \langle b_0|b_2\rangle = \langle b_1|b_2\rangle = 0.$ Alice can choose any basis.

Thus, Formula $(\ref{B2})$ always holds for all set mentioning three orthogonal product states.
Therefore, any three orthogonal product states are always distinguishable.
$\square$\\

Generally, for one-way LOCC, if Alice and Bob can decide who does the first quantum operation, we can prove the following corollary.

{\bf Corollary 3}
\emph{Any four orthogonal product states are one-way LOCC distinguishable whoever goes first.}
\begin{figure}[htp]
\centering
\includegraphics[scale=0.8]{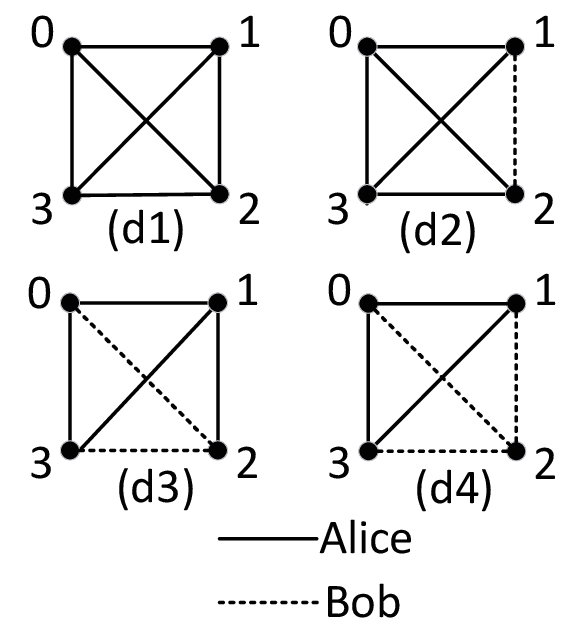}
\caption{The structure of at least three orthogonal states in Alice's or Bob's subsystems.\label{C1}}
\end{figure}

\emph{Proof:} Suppose $|\psi_k\rangle=|a_k\rangle_A|b_k\rangle_B, ~k=0,1,2,3$. In FIGs. \ref{C1} and \ref{C2}, we present the relationship of $a_k$ and $b_k$. The two points $k$ and $j$ are linked by solid lines if $|a_k\rangle$ and $|a_j\rangle $ are orthogonal in Alice's subsystems, as well as the dotted lines for Bob's subsystems.

If $\{|a_k\rangle\}$ (See FIG. \ref{C1}(d1)) or three of $\{|a_0\rangle,~|a_1\rangle,~|a_2\rangle,~|a_3\rangle\}$ (See FIG. \ref{C1}(d2, d3, d4)) are orthogonal to each other, Formula $(\ref{B2})$ holds. That is, Alice can always commence a one-way LOCC protocol to distinguish these four orthogonal product states.

Similarly for Bob, if three of $\{|b_0\rangle,~|b_1\rangle,~|b_2\rangle,~|b_3\rangle \}$ are orthogonal to each other, Formula $(\ref{B2})$ holds.

In the following, we check the other cases, which is equivalent to FIG. \ref{C2}. That is, only two equivalent classes don't have solid triangles or dotted triangles in $4$ points perfect picture as follows:

\begin{figure}[htp]
  \centering
  \includegraphics[width=0.35\textwidth]{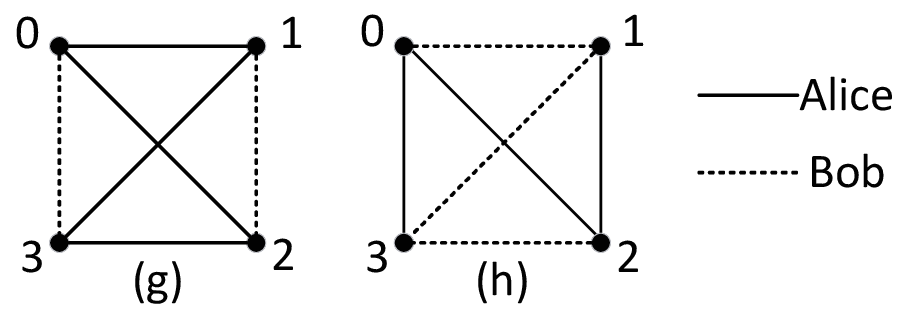}
  \caption{The structure of only two orthogonal states in Alice's or Bob's subsystems.\label{C2}}
\end{figure}

In FIG. \ref{C2}(g), we have $\langle a_0|a_1\rangle=\langle a_0|a_2\rangle=\langle a_1|a_3\rangle=\langle a_2|a_3\rangle=0$ in Alice's side and $\langle b_0|b_3\rangle=\langle b_1|b_2\rangle=0$ in Bob's side. Alice constructs the basis with $\{|a_0\rangle, |a_1\rangle\}$ via Gram-Schimdt orthogonalization to obtain
\begin{eqnarray*}
\begin{array}{l}
\displaystyle |\psi_0\rangle=|a_0\rangle_A|b_0\rangle_B, \ |\psi_1\rangle=|a_1\rangle_A|b_1\rangle_B,\\
\displaystyle |\psi_2\rangle=\alpha|a_1\rangle_A|b_2\rangle_B+(|a_2\rangle-\alpha|a_1\rangle)_A|b_2\rangle_B,\\
\displaystyle |\psi_3\rangle=\beta|a_0\rangle_A|b_3\rangle_B+(|a_3\rangle-\beta|a_0\rangle_A|b_3\rangle_B,
\end{array}
\end{eqnarray*}
where $\langle a_0|(|a_2\rangle-\alpha|a_0\rangle)=0$, $\langle a_1|(|a_3\rangle-\beta|a_1\rangle)=0$. After orthogonal normalization, these bases are $|\bar{a}_0\rangle=\frac{|a_0\rangle}{|||a_0\rangle||}$, $|\bar{a}_1\rangle=\frac{|a_1\rangle}{|||a_1\rangle||}$, $|\bar{a}_2\rangle=\frac{|a_2\rangle-\alpha|a_1\rangle}{|||a_2\rangle-\alpha|a_1\rangle||}$ and $|\bar{a}_3\rangle=\frac{|a_3\rangle-\beta|a_0\rangle}{|||a_3\rangle-\beta|a_0\rangle||}$. Formula $(\ref{B2})$ holds.

In FIG. \ref{C2}(h), we have $\langle a_0|a_2\rangle=\langle a_0|a_3\rangle=\langle a_1|a_2\rangle=0$ in Alice's side and $\langle b_0|b_1\rangle=\langle b_1|b_3\rangle=\langle b_2|b_3\rangle=0$ in Bob's side. Alice constructs the basis with Gram-Schimdt orthogonalization from $\{|a_0\rangle, |a_2\rangle\}$ to get
\begin{eqnarray*}
\begin{array}{l}
\displaystyle |\psi_0\rangle=|a_0\rangle_A|b_0\rangle_B,\ |\psi_2\rangle=|a_2\rangle_A|b_2\rangle_B,\\
\displaystyle |\psi_1\rangle=\alpha|a_0\rangle_A|b_1\rangle_B+(|a_1\rangle-\alpha|a_0\rangle)_A|b_1\rangle_B,\\
\displaystyle |\psi_3\rangle=(\beta|\bar{a}_1\rangle+\gamma|a_2\rangle)_A|b_3\rangle_B+(|a_3\rangle-\beta|\bar{a}_1\rangle-\gamma|a_2\rangle)_A|b_3\rangle_B,
\end{array}
\end{eqnarray*}
where $\langle a_0|(|a_1\rangle-\alpha|a_0\rangle)=0$, $(\beta^*\langle \bar{a}_1|+\gamma^*\langle a_2|)(|a_3\rangle-\beta|\bar{a}_1\rangle-\gamma|a_2\rangle)=0$. After orthogonal normalization, these bases are
$|\bar{a}_0\rangle=\frac{|a_0\rangle}{|||a_0\rangle||}$, $|\bar{a}_1\rangle=\frac{|a_1\rangle-\alpha|a_0\rangle}{|||a_1\rangle-\alpha|a_0\rangle||}$, $|\bar{a}_2\rangle=\frac{|a_2\rangle}{|||a_2\rangle||}$ and $|\bar{a}_3\rangle=\frac{|a_3\rangle-\beta|\bar{a}_1\rangle-\gamma|a_2\rangle}{|||a_3\rangle-\beta|\bar{a}_1\rangle-\gamma|a_2\rangle||}$. Formula (\ref{B2}) holds.

Thus, any four orthogonal states are always distinguishable via one-way LOCC no matter who goes first. $\square$
\\

According to Theorem \ref{A3}, we can estimate the number of product states of a one-way LOCC distinguishable set in $C^{d_A}\otimes C^{d_B}$ system when Alice goes first. These results are Corollary 4 and Corollary 5.

\vspace{2mm}
{\bf Corollary 4}
\emph{In a $C^{d_A}\otimes C^{d_B}$ composite system, $[\frac{d_A d_B}{2}]+k$ orthogonal states expressed as the Eq. (\ref{B2}) can be exactly one-way locally distinguished, only if at least $2k$ (when $d_Ad_B$ is even) or $2k-1$ (when $d_Ad_B$ is odd) of those states are product states. Here, ``[ $ \cdot $ ]'' is the integer-valued function.}

\vspace{2mm}
\emph{Proof:} According to Theorem \ref{A3}, any $[\frac{d_A d_B}{2}]+k$ orthogonal states expressed as $|\psi_k\rangle=\sum\limits_{j=1}^{d_A} |j\rangle_A|\eta_j^k\rangle_B,$ where $h \neq k, \langle\eta_j^h|\eta_j^k\rangle=0.$ If the set is one-way locally distinguishable, Alice firstly measures her particles A with the basis $\{|1\rangle,|2\rangle,\cdots,|d_A\rangle\}_A$ such that Bob's quantum states collapse. Without loss of generality, suppose some $|\eta_{j}^k\rangle=0$. After considering those nonzero $|\eta_{j}^k\rangle$, the number of the nonzero pair $|j\rangle_A |\eta_{j}^k\rangle_B$ is at most $d_A d_B$.

Suppose the number of product states in the set is $x$, then the number of non-product states is $[\frac{d_A d_B}{2}]+k-x$. There are at least two $|j\rangle_A |\eta_{j}^k\rangle_B$ in one non-product state and at least one $|j\rangle_A |\eta_{j}^k\rangle_B$ in one product state. Combined with Drawer Principle, therefore, $1\cdot x+2\cdot([\frac{d_A d_B}{2}]+k-x)\leq d_A d_B$ holds, i.e. $ x \geq 2k + 2\cdot[\frac{d_A d_B}{2}] - d_A d_B.$
Therefore, we obtain this corollary.$\square$

Immediately obtain the following corollary via the above one.

{\bf Corollary 5}
\emph{In a $C^{d_A}\otimes C^{d_B}$ composite system, $d_Ad_B$ orthogonal states can be exactly one-way locally distinguished if and only if all of them are product states.}

\section{The minimum structures of one-way LOCC indistinguishable pure states}
\label{sec:one}
In this section, we mainly study the minimum structures of one-way local indistinguishability of quantum pure states. We present an observation which makes it unnecessary for us to check the local indistinguishability of a whole set but only need to check a locally indistinguishable subset.

\vspace{2mm}
{\bf Observation 1}
\emph{In a $C^{d_A}\otimes C^{d_B}$ quantum system, any subset of a locally distinguishable set is locally distinguishable. It is true that there always exists at least one locally indistinguishable subset in a locally indistinguishable set.}

Firstly, let us consider the general case: the entangled state sets. Any two orthogonal pure states are always distinguishable via LOCC \cite{Walgate00}. Now, let us show the proof of the local indistinguishability of any three Bell states by using Theorem \ref{A3}. Note that,  Ghosh \emph{et al.} \cite{Ghosh01} gave the similar conclusion: any three Bell states cannot be perfectly LOCC distinguished.

{\bf Corollary 6} \emph{Any three Bell states cannot be perfectly LOCC distinguished no matter who goes first.}

\vspace{2mm}
\emph{Proof:}
Suppose three Bell states are
\begin{eqnarray}
\begin{array}{l}
\displaystyle |\Phi_0\rangle=1/\sqrt{2}(|00\rangle+|11\rangle)_{AB},\\
\displaystyle |\Phi_1\rangle=1/\sqrt{2}(|00\rangle-|11\rangle)_{AB}, \\
\displaystyle |\Phi_2\rangle=1/\sqrt{2}(|01\rangle+|10\rangle)_{AB}.
\end{array}
\end{eqnarray}
Let $|\varphi\rangle=cos\theta |0\rangle+e^{i\delta}sin\theta |1\rangle$, $|\varphi^\perp\rangle=-e^{-i\delta}sin\theta |0\rangle+cos\theta |1\rangle$, where $\theta, \ \delta\in[0,2\pi]$. We have $|0\rangle=cos\theta |\varphi\rangle-e^{i\delta}sin\theta |\varphi^\perp\rangle$, $|1\rangle=e^{-i\delta}(sin\theta |\varphi\rangle+cos\theta |\varphi^\perp\rangle.$ Therefore, we rewrite the three Bell states as follows:
\begin{eqnarray*}
\begin{array}{l}
\displaystyle |\Phi_0\rangle=1/\sqrt{2}\{|\varphi\rangle[(cos^2\theta+e^{-2i\delta}sin^2\theta)|\varphi\rangle+(e^{-i\delta}-e^{i\delta})sin\theta cos\theta|\varphi^\perp\rangle]\\
\displaystyle \qquad\quad +|\varphi^\perp\rangle[(e^{-i\delta}-e^{i\delta})sin\theta cos\theta|\varphi\rangle
+(cos^2\theta+e^{2i\delta}sin^2\theta)|\varphi^\perp\rangle]\},\\
\displaystyle |\Phi_1\rangle=1/\sqrt{2}\{|\varphi\rangle[(cos^2\theta-e^{-2i\delta}sin^2\theta)|\varphi\rangle-(e^{-i\delta}+e^{i\delta})sin\theta cos\theta|\varphi^\perp\rangle]\\
\displaystyle \qquad\quad -|\varphi^\perp[(e^{-i\delta}+e^{i\delta})sin\theta cos\theta\rangle|\varphi\rangle
+(e^{2i\delta}sin^2\theta-cos^2\theta)|\varphi^\perp\rangle]\},\\
\displaystyle |\Phi_2\rangle=1/\sqrt{2}[|\varphi\rangle (e^{-i\delta}sin2\theta|\varphi\rangle+cos2\theta|\varphi^\perp\rangle)+|\varphi^\perp\rangle (cos2\theta|\varphi\rangle\\
\displaystyle \qquad\quad -e^{i\delta}sin2\theta|\varphi^\perp\rangle)].
\end{array}
\end{eqnarray*}
\indent From the equation, we can get
\begin{eqnarray*}
\begin{array}{l}
\displaystyle|\eta_\varphi^0\rangle=(cos^2\theta+e^{-2i\delta}sin^2\theta)|\varphi\rangle+(e^{-i\delta}-e^{i\delta})sin\theta cos\theta|\varphi^\perp\rangle,\\
\displaystyle|\eta_\varphi^1\rangle=(cos^2\theta-e^{-2i\delta}sin^2\theta)|\varphi\rangle-(e^{-i\delta}+e^{i\delta})sin\theta cos\theta|\varphi^\perp\rangle,\\
\displaystyle |\eta_\varphi^2\rangle=e^{-i\delta}sin2\theta|\varphi\rangle+cos2\theta|\varphi^\perp\rangle
\end{array}
\end{eqnarray*}
and
\begin{eqnarray*}
\begin{array}{l}
\displaystyle |\eta_{\varphi^\perp}^0\rangle=(e^{-i\delta}-e^{i\delta})sin\theta cos\theta|\varphi\rangle
+(cos^2\theta+e^{2i\delta}sin^2\theta)|\varphi^\perp\rangle,\\
\displaystyle |\eta_{\varphi^\perp}^1\rangle=-(e^{-i\delta}+e^{i\delta})sin\theta cos\theta|\varphi\rangle
-(e^{2i\delta}sin^2\theta-cos^2\theta)|\varphi^\perp\rangle,\\
\displaystyle |\eta_{\varphi^\perp}^2\rangle=cos2\theta|\varphi\rangle-e^{i\delta}sin2\theta|\varphi^\perp\rangle.
\end{array}
\end{eqnarray*}
\\
\indent Three states $|\Phi_0\rangle, |\Phi_1\rangle, |\Phi_2\rangle$ are all entangled states, so none of $|\eta_\varphi^k\rangle$ and $|\eta^k_{\varphi^\perp}\rangle$ ($k=0, 1, 2$) is zero for $\forall \theta, \delta\in[0, 2\pi]$ in Bob's subsystems. If this set is to be locally distinguishable with Alice going first, there must be some choice of $\{|\varphi\rangle, |\varphi^\perp\rangle\}$ such that $\langle \eta_\varphi^0|\eta_\varphi^1\rangle=\langle \eta_\varphi^1|\eta_\varphi^2\rangle=\langle \eta_\varphi^0|\eta_\varphi^2\rangle=0$ and $\langle \eta_{\varphi^\perp}^0|\eta_{\varphi^\perp}^1\rangle=\langle \eta_{\varphi^\perp}^1|\eta_{\varphi^\perp}^2\rangle=\langle \eta_{\varphi^\perp}^0|\eta_{\varphi^\perp}^2\rangle=0$. However,
there is no room in Bob's two-dimensional Hilbert space for three mutually orthogonal states. It is impossible to satisfy the Formula (\ref{B2}), thus the set is LOCC indistinguishable. $\square$\\

Secondly, let us consider the minimum structures of one-way LOCC indistinguishable orthogonal product states.
To prove the Theorem \ref{A6} more smoothly, we first introduce Theorem \ref{A5}, Corollaries 7 and 8, which are obtained from Lemma \ref{A4}.

{\bf Corollary 7}
\emph{Two states $|\psi\rangle$ and $|\phi\rangle$ are orthogonal in $C^d$ system. The measurement $\{M_{k}\}$ can be used to distinguish $|\psi\rangle$ and $|\phi\rangle$, if and only if $\forall k,~ \langle\psi|M_{k}^\dagger M_{k}|\phi\rangle=0.$}

\begin{theorem}\label{A5}
Two states $|\psi\rangle$ and $|\phi\rangle$ are orthogonal in $C^d$ system. The rank one POVM $\{|\bar{k}\rangle\langle\bar{k}|:k=0,\cdots,d-1 \}$ can be used to distinguish $|\psi\rangle$ and $|\phi\rangle$, if and only if $\forall k,~ \langle\psi|\bar{k}\rangle\langle\bar{k}|\phi\rangle=0.$
\end{theorem}

This theorem implies the following one directly.

{\bf Corollary 8}
\emph{Two states $|\psi\rangle$ and $|\phi\rangle$ are orthogonal in $C^3$ system. The rank one POVM $\{|\bar{k}\rangle\langle\bar{k}|:k=0,1,2\}$ can be used to distinguish $|\psi\rangle $ and $ |\phi\rangle$, if and only if either $|\psi\rangle $ or $|\phi\rangle$ is in $\{|\bar{k}\rangle:k=0,1,2\}.$}

\emph{Proof:} Suppose $|\psi\rangle=|0\rangle$ and $|\phi\rangle=|1\rangle.$ Notice that in $C^3$ system, the rank one POVM $\{|\bar{k} \rangle \langle \bar{k}|: k=0,1,2 \}$ can only be one of these forms by Theorem \ref{A5}: $\{|0\rangle\langle0|, |1\rangle\langle1|, |2\rangle\langle2|\}$, $\{|0\rangle\langle0|, (\alpha|1\rangle+\beta|2\rangle)(\alpha^*\langle1|+\beta^* \langle2|), (\beta|1\rangle-\alpha|2\rangle)( \beta^*\langle1|- \alpha^*\langle2|):\alpha\beta\neq0\},$ or $\{|1\rangle\langle1|, (\alpha|0\rangle+\beta|2\rangle)(\alpha^*\langle0|+\beta^*\langle2|), (\beta|0\rangle-\alpha|2\rangle)(\beta^*\langle0|-\alpha^*\langle2|):\alpha\beta\neq0\}.$ $\square$

\begin{figure}[htp]
  \centering
  \includegraphics[scale=0.65]{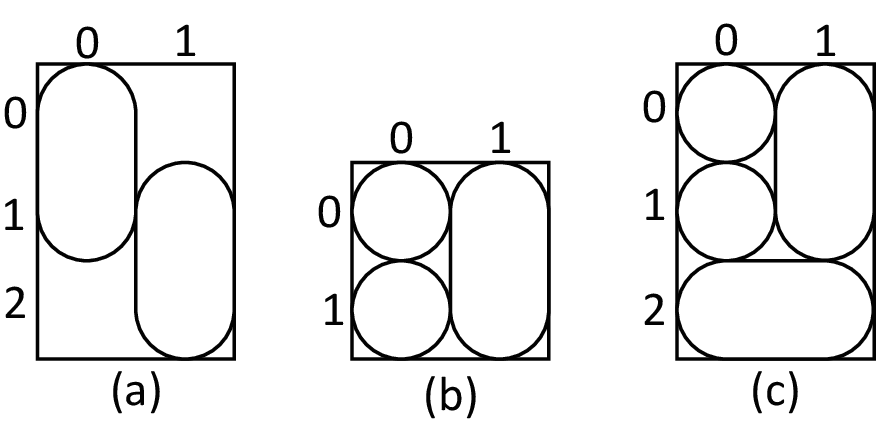}
  \caption{\label{C3} The tiling structures of orthogonal product states in $C^2\otimes C^2$ and $C^3\otimes C^2$ dimension respectively.}
\end{figure}

\begin{theorem}\label{A6}
It is four that is the least number of indistinguishable orthogonal product states via one-way LOCC when Alice goes first.
\end{theorem}
\emph{Proof:}
In FIG. \ref{C3}(a), these states are as follows:
$$|\psi_{1,2}\rangle=\frac{1}{\sqrt{2}}(|0\rangle\pm|1\rangle)_A|0\rangle_B,\
|\psi_{3,4}\rangle=\frac{1}{\sqrt{2}}(|1\rangle\pm|2\rangle)_A|1\rangle_B.$$

For this set of orthogonal product states, Bob cannot distinguish $|\psi_{1}\rangle$ and $|\psi_{2}\rangle$ (or $|\psi_{3}\rangle$ and $|\psi_{4}\rangle$) in his subsystems. By Theorem \ref{A3} and Corollary 8, if these four states are distinguishable, one of $\{ \frac{1}{\sqrt{2}}(|0\rangle+|1\rangle)_A$, $\frac{1}{\sqrt{2}}(|0\rangle-|1\rangle)_A \}$ must be chosen by Alice to distinguish $|\psi_{1}\rangle $ and $|\psi_{2}\rangle$ , also for $\frac{1}{\sqrt{2}}(|1\rangle+|2\rangle)_A$ and $\frac{1}{\sqrt{2}}(|1\rangle-|2\rangle)_A$.
However, $\frac{1}{\sqrt{2}}(|0\rangle+|1\rangle)_A$ and $\frac{1}{\sqrt{2}}(|1\rangle\pm|2\rangle)_A$ are not orthogonal.
$\frac{1}{\sqrt{2}}(|0\rangle-|1\rangle_A$ and $\frac{1}{\sqrt{2}}(|1\rangle\pm|2\rangle)_A$ are not orthogonal either. This means these four states are indistinguishable via one-way LOCC when Alice goes first.

Similarly in FIG. \ref{C3}(b), the set $\{|\phi_0\rangle=|0\rangle_A|0\rangle_B,|\phi_1\rangle=|1\rangle_A|0\rangle_B,
|\phi_{2,3}\rangle=\frac{1}{\sqrt{2}}(|0\rangle\pm|1\rangle)_A|1\rangle_B\}$ is also indistinguishable via one-way LOCC when Alice goes first. In fact, Groisman and Vaidman have shown the one-way LOCC indistinguishability for the above set by a different method \cite{Groisman01}. $\square$\\

With above conclusions, we can also judge immediately the one-way local indistinguishability of orthogonal product states in Eq. (\ref{B1}). We choose these states
$|\psi_{1,2}\rangle=|0\rangle_A(|\frac{1}{2}|0\rangle\pm\frac{1}{\sqrt{2}}|1\rangle+\frac{1}{2}|2\rangle)_B,$ $|\psi_{3,4}\rangle=(|\frac{1}{2}|0\rangle\pm\frac{1}{\sqrt{2}}|1\rangle+\frac{1}{2}|2\rangle)_A|3l_B\rangle_B,$ $|\psi_5\rangle=|1\rangle_A|(3l_B-1)\rangle_B,$ $|\psi_6\rangle=|2\rangle_A|(3l_B-1)\rangle_B.$ The tiling structure is similar to FIG. \ref{C3}(a), thus the subset is one-way LOCC indistinguishable when Alice goes first. When Bob goes first, we can also find the similar structure. Therefore, the set in Eq. (\ref{B1}) is one-way LOCC indistinguishable no matter who goes first.

We continue to check whether there exists five indistinguishable orthogonal product states by one-way LOCC.
Notice that, DiVincenzo \emph{et al.} has found the very similar example in \cite{DiVincenzo03}. The indistinguishability of this special set can be proved by different ways.

{\bf Corollary 9}
\emph{There exist five orthogonal product indistinguishable states via one-way LOCC no matter who goes first.}

\begin{figure}[htp]
  \centering
  \includegraphics[scale=0.6]{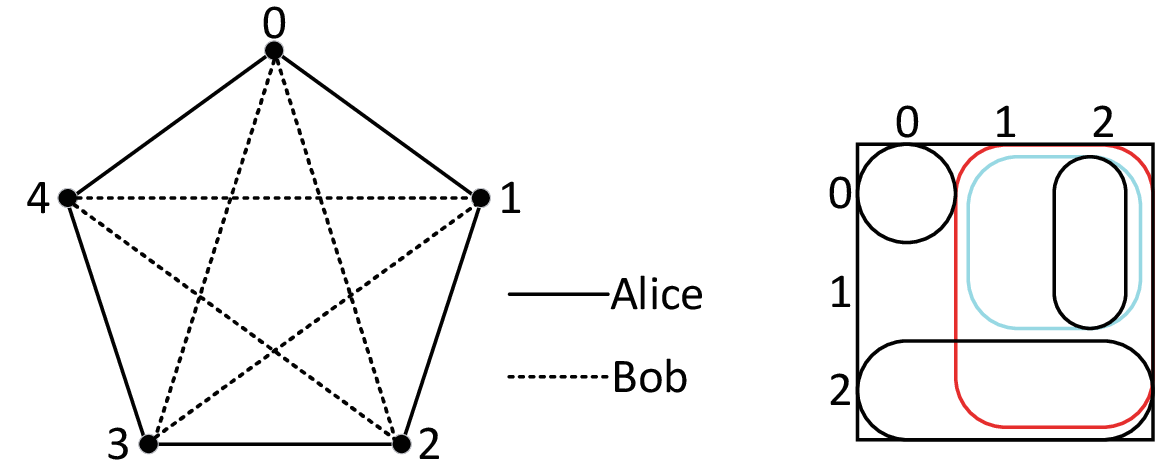}
  \caption{\label{C4} (Color online) The structure of five indistinguishable orthogonal product states by one-way LOCC.}
\end{figure}

\emph{Proof:} In FIG. \ref{C4}, suppose $|\Psi_k\rangle=|a_k\rangle_A|b_k\rangle_B$, specifically,
$|\Psi_0\rangle=|0\rangle_A|0\rangle_B,~|\Psi_1\rangle= \frac{1}{\sqrt{3}} |2\rangle_A(|0\rangle-|1\rangle+|2\rangle)_B,$
$|\Psi_2\rangle=\frac{1}{\sqrt{2}}(|0\rangle+|1\rangle)_A|2\rangle_B,~
|\Psi_3\rangle=\frac{1}{\sqrt{6}}(|0\rangle-|1\rangle+|2\rangle)_A(|1\rangle+|2\rangle)_B,$
$|\Psi_4\rangle=\frac{1}{2}(|1\rangle+|2\rangle)_A(|0\rangle+|1\rangle)_B.$

If $|a_k\rangle$ and $|a_j\rangle $ in Alice's subsystems are orthogonal then the solid lines link two points $k$ and $j$, and the dotted lines for Bob's subsystems. If the five states are one-way distinguishable by Theorem \ref{A3} when Alice goes first (similar proof when Bob starts), we check which basis should be chosen in Alice's side. For the convenience of the readers, suppose $|\Psi_5\rangle=|a_5\rangle_A|b_5\rangle_B=|a_0\rangle_A|b_0\rangle_B=|\Psi_0\rangle.$

We will use a lemma: `` $\forall k,$ either $|a_k\rangle_A $ or $|a_{k+1}\rangle_A$ of the orthogonal pair $\{|a_k\rangle_A, |a_{k+1}\rangle_A\}$ is in the chosen basis for $k=0,1,2,3,4$". This lemma can be founded because of $\langle b_{k}|b_{k+1}\rangle\neq 0$ and Corollary 7.

However, the contradiction will be produced. For instance, if we choose $|a_0\rangle$, it means that we cannot choose $|a_1\rangle$, so we must choose $|a_2\rangle.$ But $|a_0\rangle$ and $|a_2\rangle$ are not orthogonal, this is a contradiction. Otherwise, if we choose $|a_1\rangle$, it means that we cannot choose $|a_2\rangle$, so we must choose $|a_3\rangle.$ However, $\langle a_3|a_1\rangle\neq 0,$ this also produces contradiction. Therefore, the five orthogonal product states are indistinguishable via one-way LOCC. $\square$
\\

In $C^3\otimes C^2$ (See FIG. \ref{C3}(c)), we further find that six orthogonal product states are one-way LOCC indistinguishable whoever goes first. Their forms are as follows:
\begin{eqnarray}
\begin{array}{l}
\displaystyle |\Psi_1\rangle=|0\rangle_A|0\rangle_B,\ |\Psi_2\rangle=|1\rangle_A|0\rangle_B,\\
\displaystyle |\Psi_{3,4}\rangle=\frac{1}{\sqrt{2}}(|0\rangle\pm |1\rangle)_A|1\rangle_B,\\
\displaystyle |\Psi_{5,6}\rangle=|2\rangle_A\frac{1}{\sqrt{2}}(|0\rangle\pm |1\rangle)_B.
\end{array}
\end{eqnarray}

\section{Conclusion}
\label{sec:con}
We have provided sufficient and necessary conditions respectively for distinguishing bipartite pure quantum states and mixed quantum states by one-way LOCC. We give some applications to show that the distinguishability of a state set can be proved efficiently via these conditions. The indistinguishability of a state set can be checked via searching out one indistinguishable subset in it. Sometimes only four orthogonal product states or three orthogonal maximally entangled states are needed to find via our results. Moreover, we also provide a bound of the number of orthogonal product states in a distinguishable state set.

\begin{acknowledgments}
This work was supported by National Key R\&D Plan of China (Grant No. 2017YFB0802203), National Natural Science Foundation of China (Grant Nos. U1736203, 61877029, 61872153, 61732021, 61472165, 61373158, 61672014 and 61502200), Guangdong Provincial Engineering Technology Research Center on Network Security Detection and Defence (Grant No. 2014B090904067), Guangdong Provincial Special Funds for Applied Technology Research and Development and Transformation of Important Scientific and Technological Achieve (Grant No. 2016B010124009), Natural Science Foundation of Guangdong Province (Grant No. 2016A030313090), the Zhuhai Top Discipline--Information Security, Guangzhou Key Laboratory of Data Security and Privacy Preserving, Guangdong Key Laboratory of Data Security and Privacy Preserving, National Joint Engineering Research Center of Network Security Detection and Protection Technology, National Cryptography Development Fund MMJJ20180109, and the Fundamental Research Funds for the Central Universities and Guangdong Provincial Special Funds for Applied Technology Research and Development and Transformation of Important Scientific and Technological Achieve (Grant No. 2017B010124002).
\end{acknowledgments}

\nocite{*}
\bibliographystyle{elsarticle-num}
\bibliography{apssamp}
\section*{ Appendix}
Appendix is the proof of \emph{Corollary 1}.

\emph{Corollary 1.
In $C^{3l_A+1}\otimes C^{3l_B+1}$ $(1\leq l_A\leq l_B)$ quantum system, there exists an indistinguishable set via one-way LOCC no matter who goes first, which contains $4l_Al_B+7l_A+3l_B-3$ orthogonal product states $|\varphi_s\rangle$ (in Eq. (\ref{B1})).}

\vspace{2mm}
\emph{Proof:}
We only consider that Alice goes first and the same as Bob. A set of general $C^{(3l_A+1)}\otimes C^{(3l_B+1)}$ POVM elements $M_m^\dagger M_m$ under the basis $\{|0\rangle,|1\rangle,\cdots, |3l_A\rangle\}_A$ can be expressed as

$$M_m^\dagger M_m=(a_{ij}^{m}), where \ a_{ij}^{m}\geq0\ with \ i, j\in\{0, 1, 2, \cdots, 3l_A\}.$$

Firstly, we point that this selected sets $\{|0\rangle, |1\rangle, |2\rangle\}_A$, $\{|1\rangle, |2\rangle, |3\rangle\}_A$, $\cdots$, and $\{|3l_A-2\rangle, |3l_A-1\rangle, |3l_A\rangle\}_A$ of states are of dimension $C^3\otimes C^{(3l_B+1)}$, Alice cannot find appropriate basis to express them in the form of Eq.(\ref{B2}) after Alice performs measurement. Notice, the product states $|q_2\rangle|3l_B-1\rangle$ ($q_2=1, 2, 3, \cdots, 3l_A-1$) can only be distinguished where the measurements are $\{|1\rangle\langle 1|, |2\rangle\langle2|, \cdots, |3l_A-1\rangle\langle 3l_A-1|\}_A$. There no exists the superposition state of $|0\rangle$ and $|3l_A\rangle$ (See Fig. 1), so the remaining two elements in the above measurements are $|0\rangle\langle 0|$ and $|3l_A\rangle\langle 3l_A|$. The effect of this positive operator upon states
\begin{eqnarray*}
\begin{array}{l}
\displaystyle |\varphi_{i_dj_d}^\pm\rangle=|i_d\rangle_A(\frac{1}{2}|j_d\rangle\pm\frac{1}{\sqrt{2}}|j_d+1\rangle+\frac{1}{2}|j_d+2\rangle)_B, where\\
\displaystyle d=1\ and \ i_1=0,1, 2,\  j_1=i_1, i_1+3,\cdots, 3l_B-2i_1-3,\\
\displaystyle |\varphi_{i_ej_e}^\pm\rangle=(\frac{1}{2}|i_e\rangle\pm\frac{1}{\sqrt{2}}|i_e+1\rangle+\frac{1}{2}|i_e+2\rangle)_A|j_e\rangle_B,where\\
\displaystyle e=2\ and \  j_2=3l_B,\  i_2=(3l_B-j_2)/2,\\
\displaystyle |\varphi_{i_fj_f}\rangle=|i_f\rangle_A|j_f\rangle_B, where\\
\displaystyle f=2 \ and \ i_2=1, 2, j_2=3l_B-1.
\end{array}
\end{eqnarray*}
is entirely specified by those elements in the submatrix $(a_{ij}^m)$ drawn from the subspace $\{|0\rangle, |1\rangle, |2\rangle\}_A$, where $i, j\in\{0, 1, 2\}$. It means that Alice cannot perform a nontrivial measurement upon the subspace $\{|0\rangle,$ $ |1\rangle,|2\rangle\}_A$. Thus, the corresponding submatrix must be proportional to the identity. Then, we obtain $a_{00}=a_{11}=a_{22}=a,$ $ a_{01}=a_{02}=a_{10}=a_{20}=a_{12}=a_{21}=0$. For the states
\begin{eqnarray*}
\begin{array}{l}
\displaystyle |\varphi_{i_dj_d}^\pm\rangle=|i_d\rangle_A(\frac{1}{2}|j_d\rangle\pm\frac{1}{\sqrt{2}}|j_d+1\rangle+\frac{1}{2}|j_d+2\rangle)_B,where\nonumber\\
\displaystyle d=1 \ and \ i_1=1, 2, 3, \ j_1=i_1, i_1+3,\cdots, 3l_B-2i_1-3,\nonumber\\
\displaystyle |\varphi_{i_ej_e}^\pm\rangle=(\frac{1}{2}|i_e\rangle\pm\frac{1}{\sqrt{2}}|i_e+1\rangle+\frac{1}{2}|i_e+2\rangle)_A|j_e\rangle_B,where\nonumber\\
\displaystyle j_1=0, i_1=j_1+1,\nonumber\\
\displaystyle |\varphi_{i_fj_f}\rangle=|i_f\rangle_A|j_f\rangle_B, where\nonumber\\
\displaystyle f=2 \ and \ i_2=1, 2, 3,\  j_2=3l_B-1.
\end{array}
\end{eqnarray*}
and the subspace $\{|1\rangle, |2\rangle, |3\rangle\}_A$, we make the same argument. Then we obtain results $a_{11}=a_{22}=a_{33}=a,$ $ a_{12}=a_{13}=a_{21}=a_{31}=a_{23}=a_{32}=0$. In the same way, for the subspace $\{|2\rangle, |3\rangle,|4\rangle\}_A$, $\{|3\rangle, |4\rangle,|5\rangle\}_A$, $\cdots$ and the subspace $\{|3l_A-2\rangle, |3l_A-1\rangle, |3l_A\rangle\}_A$, we obtain result
\begin{eqnarray*}
&&{}a_{44}=a_{55}=\cdots=a_{3l_A,3l_A}=a, \nonumber\\
&&{}a_{45}=a_{46}=a_{56}=\cdots=a_{3l_A-1,3l_A}=a_{3l_A,3l_A-1}=0.
\end{eqnarray*}
Because the POVM elements $M_m^\dagger M_m$ is Hermitian, $(M_m^\dagger M_m)^\dagger=M_m^\dagger M_m$ holds. Then we obtain
$a^*=a, a_{30}=a_{03}^*, a_{40}=a_{04}^*,\cdots,a_{3l_A,3l_A-3}=a_{3l_A-3,3l_A}^*.$ We now consider the states
\begin{eqnarray*}
&&{}|\varphi_{i_dj_d}^\pm\rangle=|i_d\rangle_A(\frac{1}{2}|j_d\rangle\pm\frac{1}{\sqrt{2}}|j_d+1\rangle+\frac{1}{2}|j_d+2\rangle)_B,where \nonumber\\
&&{}d=1\ and \  i_1=0, \ j_1=3l_B-2i_1-3,\nonumber\\
&&{}|\varphi_{i_fj_f}\rangle=|3\rangle_A|3l_B-1\rangle_B
\end{eqnarray*}
and the subspace $\{|0\rangle, |3\rangle\}_A$. After Alice measures, the result is either the states orthogonal or distinguishing them outright. If the result is the states orthogonal, we demand that $\langle\varphi_{d=1}|M_m^+M_m|\varphi_{f=3l_B+2}\rangle=\frac{1}{2}a_{03}=0$. So, we obtain $a_{03}^*=a_{03}=0$. For the subspace $\{|0\rangle, |4\rangle\}_A$,$\cdots$ and the subspace $\{|3l_A-3\rangle, |3l_A\rangle\}_A$, we can obtain results
\begin{eqnarray*}
&&{}a_{04}=a_{04}^*=a_{05}=a_{05}^*=\cdots=a_{3l_A,3l_A-3}=a_{3l_A-3,3l_A}^*=0.
\end{eqnarray*}
Now the $M_m^\dagger M_m$ is proportional to the identity.

However, if Alice distinguishes the state
\begin{eqnarray*}
&&{}|\varphi_{i_dj_d}^\pm\rangle=|i_d\rangle_A(\frac{1}{2}|j_d\rangle\pm\frac{1}{\sqrt{2}}|j_d+1\rangle+\frac{1}{2}|j_d+2\rangle)_B, where\nonumber\\
&&{} d=1\ and \ i_1=0,\  j_1=3l_B-2i_1-3,\nonumber\\
&&{}|\varphi_{i_fj_f}\rangle=|3\rangle_A|3l_B-1\rangle_B,
\end{eqnarray*}
we get the result $\langle\varphi_i|M_m^\dagger M_m|\varphi_i\rangle=0$. We can also obtain the result $\langle\varphi_i|M_m^\dagger M_m|\varphi_i\rangle=\frac{a}{2}$, therefore $a=0$. It contradicts the Theorem \ref{A3} in this paper. So, $M_m^\dagger M_m$ is proportional to the identity and the orthogonal product states are indistinguishable.$\square$

\end{document}